# Structural and optical properties of single- and few-layer magnetic semiconductor CrPS$_4$

Jinhwan Lee[1†], Taeg Yeoung Ko[2†], Jung Hwa Kim[3], Hunyoung Bark[4], Byunggil Kang[4], Soon-Gil Jung[5,6], Tuson Park[5,6], Zonghoon Lee[3], Sunmin Ryu[2,7*], and Changgu Lee[1,4*]

[1]School of Mechanical Engineering, Sungkyunkwan University, 2066 Seobu-ro, Jangan-gu, Suwon, Gyunggi-do 16419, South Korea

[2]Department of Chemistry, Pohang University of Science and Technology (POSTECH), 50 Jigokro 127, Pohang, Gyeongbuk 37673, South Korea

[3]School of Materials Science and Engineering, Ulsan National Institute of Science and Technology (UNIST), 50 UNIST-gil, Eonyang-eup, Ulsan 44919, South Korea

[4]SKKU Advanced Institute of Nanotechnology (SAINT), Sungkyunkwan University, 2066 Seobu-ro, Jangan-gu, Suwon, Gyunggi-do 16419, South Korea

[5]Center for Quantum Materials and Superconductivity (CQMS), Sungkyunkwan University, 2066 Seobu-ro, Jangan-gu, Suwon, Gyunggi-do 16419, South Korea

[6]Department of Physics, Sungkyunkwan University, 2066 Seobu-ro, Jangan-gu, Suwon, Gyunggi-do 16419, South Korea

[7]Division of Advanced Materials Science, Pohang University of Science and Technology (POSTECH), 50 Jigokro 127, Pohang, Gyeongbuk 37673, South Korea

ABSTRACT

Atomically thin binary 2-dimensional (2D) semiconductors exhibit diverse physical properties depending on their composition, structure and thickness. By adding another element in those materials, which will lead to formation of ternary 2D materials, the property and structure would greatly change and significantly expanded applications could be explored. In this work, we report structural and optical properties of atomically thin chromium thiophosphate ($CrPS_4$), a ternary antiferromagnetic semiconductor. Its structural details were revealed by X-ray and electron diffractions. Transmission electron microscopy showed that preferentially-cleaved edges are parallel to diagonal Cr atom rows, which readily identified their crystallographic orientations. Strong in-plane optical anisotropy induced birefringence that also enabled efficient determination of crystallographic orientation using polarized microscopy. The lattice vibrations were probed by Raman spectroscopy for the first time and exhibited significant dependence on thickness of crystals exfoliated down to single layer. Optical absorption determined by reflectance contrast was dominated by d-d type transitions localized at $Cr^{3+}$ ions, which was also responsible for the major photoluminescence peak at 1.31 eV. The spectral features in the absorption and emission spectra exhibited noticeable thickness-dependence and hinted a high photochemical activity for single layer $CrPS_4$. The current structural and optical investigation will provide a firm basis for future study and application of this novel magnetic semiconductor.

The emergence of two-dimensional (2D) materials, such as graphene, have attracted enormous attention in the past decade because of their remarkable mechanical, electrical and physical properties.[1-4] Although graphene has attractive intrinsic characteristics such as high carrier mobility and low electrode-channel contact resistance, it is not useful for digital switching applications requiring sizable electronic bandgap owing to its zero-bandgap.[5-7] Hence, layered transition metal chalcogenides (TMCs) with bandgap ranging from 0.5 – 2.0 eV have emerged as good candidate materials for next generation digital electronics. So far, binary TMCs (BTMCs), such as $MoS_2$, $WSe_2$, $MoTe_2$, $ReS_2$, and SnS, have been intensively studied and their electrical and optical properties were characterized for electronic and optoelectronic applications.[4, 8-10] However, these BTMCs mostly lack magnetism, thus are expected to have limited applications in conventional types of electronic devices.[11] In contrast, ternary transition metal chalcogenides (TTMCs), which consist of three elements, can have much more combinations and diverse physical properties. Some TTMCs with magnetic properties have been widely studied since the 1970s for their growth, structure, magnetism, ion intercalation, and so on.[12-14]

Several TTMCs, in the form of $MAX_n$ (M= transition metal such as Mn, Fe, Ni, Cr, Co, A=P, Si, Ge, and X=S, Se, Te, n=3, 4), exhibit ferromagnetism (FM) or antiferromagnetism (AFM) that originates from transition metal ions. When their thickness is reduced down to single layer (1L), their magnetic and electronic properties are expected to vary as was witnessed in many BTMCs.[15-17] For example, 1L $MnPSe_3$ is predicted to maintain antiferromagnetism like its bulk form at the ground state. When it is electron- or hole-doped, however, it turns into a ferromagnetic half-metal and its spin-polarization depends on the type of charge carriers.[18] For 1L $CrSiTe_3$, which exhibits ferromagnetic ordering below 32 K in its bulk form, the Curie temperature was predicted to increase to 80 K.[16] Another theoretical study predicted that the Curie temperature of 1L $CrSiTe_3$ can be increased to near room temperature owing to enhanced ferromagnetic ordering by applying 8% strain.[19] Since the interlayer cohesion energy of several TTMCs were predicted to be smaller than that of graphite,[17] 1L TTMCs can be readily prepared by the mechanical exfoliation method invented for graphene.[2] While magnetic phase transition of single and few-layer $FePS_3$ has recently been investigated using Raman spectroscopy,[20, 21] most of 2D TTMCs remain unexplored despite the aforementioned rich magnetic properties as well as the semiconducting characteristics, which is suitable for spintronic device applications adding another dimension to the conventional electronics.

Among various TTMCs, chromium thiophosphate ($CrPS_4$) is a promising antiferromagnetic material with a Neel temperature of 36 K in a bulk form[13], but have been reported in only few research papers since 1977. It exhibits an anisotropic magnetization behavior between *ab* plane and *c* axis under the Neel temperature. A recent paper predicted that A-type AFM state is energetically most favorable,[22] which suggests intriguing FM-AFM alternation for odd and even-numbered layers. $CrPS_4$ also draws more attention due to its semiconducting characteristics as revealed by a temperature dependence of resistivity and optical bandgap by the fundamental absorption edge at 1.4 eV.[13, 23] As such in $CrPS_4$, $Cr^{3+}$ ions doped in various host materials exhibited rich photophysics and have been used in various photonic applications as represented by the first laser using ruby as a gain medium.[24] As a layered material, intercalation behavior has been studied for $Li_xCrPS_4$ in a perspective of possible use in batteries.[13] While X-ray absorption spectroscopy has shown that the empty band structures are similar to those of $MPX_3$,[25] $CrPS_4$ has an asymmetric in-plane structure and thus can have anisotropic material properties depending on the crystalline orientation. Lack of inversion symmetry should endow this material with additional functionality such as piezoelectricity and optical second-harmonic generation along with magnetic and semiconducting behaviors.[26] Moreover, the thickness may be exploited to a new degree of freedom to vary the material properties of 2D CrPS$_4$. Despite these possibilities, however, current understanding on bulk CrPS$_4$ is very limited and no investigation has been made for 2D CrPS$_4$.

In this work, we synthesized high-quality CrPS$_4$ crystals using chemical vapor transport (CVT) and performed optical studies on mechanically isolated 2D CrPS$_4$ of thickness ranging from single to hundreds of layers. Structural analyses using X-ray diffraction (XRD) revealed its space group $C_2^3$ and lattice constants. The crystalline orientation of the exfoliated 2D samples which often cleave with preferential edge directions were determined with high-resolution transmission electron microscopy (TEM) combined with polarized optical microscopy (OM). Thickness-dependent lattice vibrations were revealed by Raman spectroscopy and also exploited as a thickness probe along with an atomic force microscope (AFM). Photoluminescence (PL) and differential reflectance spectroscopy showed that the d-d transitions of Cr ions are mostly responsible for the optical transitions. The observed weak dependence on the thickness was attributed to the strong localization on the meal-ligand bonds. Nonetheless, 1L CrPS$_4$ exhibited an unusually high susceptibility towards photo and ambient oxidation, which demands a systematic investigation for future research and applications.

**RESULTS AND DISCUSSION**

**Synthesis and characterizations of single crystals.** The crystal structure of CrPS$_4$ has monoclinic symmetry with $a$= 10.871 Å, $b$= 7.254 Å, $c$= 6.140 Å, $\beta$= 91.88°, volume= 483.929 Å$^3$, space group $C_2^3$ and Z= 4.[27] As shown in Figure 1a, the sulfur atoms are regularly lying by puckered layers arranged in hexagonal close-packing on (001) face. These six sulfur atoms form a slightly distorted octahedron, where a chromium atom is coordinated in the middle. The phosphorus atoms are located in the center of the tetrahedron, which consists of four sulfurs. Because these tetrahedrons are located between columns of the octahedrons along the *b* axis, distances between neighboring Cr atoms are longer along the *a* axis than *b*, which creates a structural in-plane anisotropy.[27] In the view of *ac* plane, $CrPS_4$ belongs to the category of layered materials since it has weak van der Waals gaps between sulfur layers. Thus, thin flakes could be easily isolated using the traditional mechanical cleavage technique well known as the Scotch tape method.[13, 28]

The single crystal flakes of $CrPS_4$ could be synthesized by CVT method.[13] The mixture of powdered elements of Cr, P, and S with a rough atomic ratio of 1:1:4 was annealed at 700 ℃ for a week in a vacuum-sealed ampule, and a few millimeter-sized shiny flakes were formed (see Figure S1 in supplementary information). Its composition ratio was measured by the energy dispersive spectroscopy (EDS) and was close to 1:1:4 for Cr:P:S, which matches natural atomic ratio of $CrPS_4$ (see Figure S2). The Figure 1b shows AFM image of $CrPS_4$ exfoliated on $SiO_2$/Si substrate by the Scotch tape method. The inset shows an optical microscope image of the AFM scan area marked with a dashed white box. Thin layers including single layer could be identified easily with the optical microscope, and the thickness of single layer was about 1 nm according to AFM topography measurement. Although the interlayer spacing is about 0.62 nm as evidenced by TEM cross-sectional imaging (Figure S3), the thickness of a single layer of 2D materials is usually measured to be slightly larger in AFM topography than the interlayer spacing due to the van der Waals interaction difference between the layer and the substrate.[29, 30] However, the height of steps across similar materials is multiples of the interlayer spacing. As shown in the AFM image, the single layer $CrPS_4$ was quite flat and stable in the air, which is advantageous in handling the material for its characterization.

The magnetic property measurement system (MPMS), which has a sensitive superconducting quantum interference device (SQUID) magnetometer, was used to characterize the magnetic property of the synthesized bulk $CrPS_4$ crystal. The temperature-dependent magnetization measurement revealed that the synthesized sample has antiferromagnetism with the Neel temperature of 37.9K (see Figure S4). Furthermore, the electronic transport measurements from a field effect transistor made of a bulk crystal showed n-type semiconducting behavior, and thus electrons are the main charge carrier in this material. The measured mobility was relatively low ($1.0 \times 10^{-4}$ $cm^2/Vs$) with poor on-current ($10^{-10}$ A) for thickness of 470 nm. Even though the measured bandgap is not so wide according to our PL measurements (to be presented below), the electrical conductivity at on-state is quite low compared to other semiconductors and needs to be investigated in depth in the future (see Figure S5).

**Structural analysis by electron probe.** The high-resolution scanning TEM (STEM) image in Figure 2a precisely revealed the periodic atomic arrangement of the synthesized $CrPS_4$. In z-contrast STEM image, the brightest white spot represents Cr columns. The distance between Cr columns is around 0.54 and 0.36 nm in *a* and *b* axis, which correspond to the distance between Cr atoms. As shown in the atomic model in Figure 2a, the dashed rectangle box designates a unit cell of $CrPS_4$ in (001) face, which is also denoted with the red and green box in Figure 1a. The X-ray diffraction (XRD) method was employed to study the crystal structure of the synthesized sample as shown in Figure S6. The result coincided well with the calculated pattern, which was obtained from Mercury software. Also, the SAED results as shown in the inset of Figure 2a revealed single crystallinity clearly and were consistent with the d-spacing values from the XRD results (see Table S1 in supplementary information).

Viewing the atomic crystal structure in Figure 2, CrPS$_4$ is structurally anisotropic along the basal plane owing to the different distances between Cr atoms in *a* and *b* axes (see Figure S7). This anisotropy may induce unique phenomena such as crystallographic orientation-dependent optical and electrical properties, and piezoelectricity. Also, anisotropic mechanical properties may lead to cleavage of thin crystals along preferred crystallographic directions.[31-36] As expected, exfoliated CrPS$_4$ samples were terminated with straight edges that formed characteristic angles as shown in Figure 2b. Since the edge preference originates from overall energy gain when cleaved along certain crystallographic directions, the two characteristic angles could be related to certain angles between crystallographic directions. Hence, its crystal orientation can be identified by comparing the angles in the OM image and those in the STEM image. Among many possibilities in *ab* plane, there were two angles, 67.5 and 112.5 degrees as shown in Figure 2a, which agreed well with the two characteristic angles found in the OM image of Figure 2b. By comparing the lattice structure model overlaid on the STEM and OM images, we found that the preferred edges are aligned along the Cr diagonal directions (dashed white line in Figure 2b). High resolution STEM measurements carried out for straight edges of exfoliated thin CrPS$_4$ directly confirmed this atomic alignment as shown in Figure S8. These observations show that CrPS$_4$ crystals tend to be easily torn along the diagonal direction of a Cr atom rectangle unlike common expectation. This analysis provides an easy tool to determine the crystallographic orientation of exfoliated CrPS$_4$ and can be readily used in studying orientation-dependent physical properties.

**Polarized optical microscopy.** Among 2D materials having in-plane structural anisotropy, black phosphorus (BP) is a well-studied representative with a puckered structure.[36, 37] Its electrical conductivity and optical properties are strongly dependent on crystallographic orientation.[37, 38] CrPS$_4$, owing to the aforementioned anisotropy, is also expected to have a similar behavior. Figure

3 shows the optical response to the polarized light irradiation from an optical microscope. In this experiment, the light polarized by a linear polarizer (P) along the horizontal direction was incident on the sample with different thickness and the reflected light was detected with an analyzer (A) preset on a certain crystallographic orientation (*a* axis) as shown in Figure 3a. Then the sample was rotated clockwise by an increment of 15 degrees to see the angular dependence of the reflection. As can be seen in Figure 3b and Figure S9, the reflected light intensity was minimum at zero degree corresponding to a position of extinction and reached a maximum at 45 degrees showing further intensity modulation with a period of 90 degrees.

As a monoclinic crystal, $CrPS_4$ is optically biaxial and may exhibit birefringence. Linearly polarized light propagating perpendicular to the basal plane can be described by a vector sum of two orthogonal polarizations, characteristic of *a* and *b* axes, which will experience different refractive indices. Thus, the in-plane anisotropy in $CrPS_4$ leads to reflected light of elliptical polarization in general. For this reason, reflected light could be observed even though the polarizer and analyzer were aligned perpendicularly to each other. Also, the reflection reached a maximum when the angles between the incident polarization and the principal axes are 45 or 135 degrees, which is expected from a birefringent material.[39] This trend is more clearly identified with angle-dependent variations of red, green, and blue colors. To analyze the optical contrast quantitatively, we extracted average red, green and blue (RGB) reflection signal from six spots with increasing thickness from 50 to 90 nm as shown in the polar plots of Figure 3c (see Figure S10 for further details). The angular dependence shaped like a four-petal flower became more conspicuous with increasing thickness for all three colors, which indicating that the thicker samples reflect more light than thinner ones. For single - and few-layered samples, we could not obtain a noticeable contrast because of weak light reflection. However, it would be possible with a stronger incident light. Since some samples are not exfoliated with the preferred edges, the current polarized microscopy will serve as a powerful tool to identify crystallographic orientations of CrPS$_4$ samples with arbitrary shape.

**Thickness-dependent lattice vibrations.** Lattice vibrations of 2D crystals have revealed interesting behaviors owing to confinement along the stacking axis.[40, 41] In order to establish optical metrology for thickness and quality, we carried out systematic Raman spectroscopy measurements. Although exfoliated samples were apparently stable in the ambient conditions, photodamage was observed at a very low power density of ~10 μW/μm$^2$ (Figure S11). To avoid the photoreactions, samples were placed in an optical cell filled with flowing Ar gas for all the Raman measurements. Figure 4 presents the Raman spectra of 1 ~ 5L and bulk-like (denoted by bulk hereafter) thick samples obtained at the excitation wavelength of 514 nm. To our knowledge, the lattice vibrations of bulk CrPS$_4$ have not been studied yet and their detailed analyses will be reported as a separate paper employing polarized Raman spectroscopy. In this work, most salient features that depend on the number of layers will be emphasized. Briefly, bulk CrPS$_4$ is monoclinic and belongs to a space group of $C_2^3$ with a unit cell consisting of two formula units (Cr$_2$P$_2$S$_8$) in each layer. Assuming very weak interlayer interaction as in other ternary layered crystals,[17] the lattice vibrations of bulk CrPS$_4$ can be described as those of its single layer building block.[42] From a group theoretical analysis,[43] the lattice vibrations at the Brillouin zone center (Γ) can be decomposed into the following irreducible representations of the point group C$_2$: Γ = 17A + 19B, where all modes except acoustic modes (A+2B) are Raman-active. In a backscattering geometry with linearly-polarized excitation laser beam normally incident to the basal plane, only A (B) modes can be observed when Raman signals whose polarization is parallel (perpendicular) to that of the input are selected for detection. To characterize all the Raman peaks, both polarization components were detected without selection.

From bulk samples (Figure 4a), 17 Raman peaks were identified at the excitation wavelength of 514 nm in the range of 100 ~ 700 cm$^{-1}$ and alphabetically labeled in the order of increasing frequency. The spectra obtained at a wavelength of 633 nm exhibited two more peaks that were not observed at 458 and 514 nm (Figure S12). While the dependence on the excitation wavelength may suggest existence of an electronic resonance near 1.9 eV, absorption spectra of bulk samples in a previous work[13] and our own measurements (Figure S13) did not show such a transition. Since 33 optical phonon modes are all Raman-active, more peaks can be detected in high-sensitivity measurements that are currently impeded by photodegradation. In the Raman spectra of 2D CrPS$_4$ (Figure 4a), most of the bulk Raman peaks were found except 1L. As will be corroborated by the optical reflectance and PL data below, 1L did not generate detectable Raman signal possibly because of its photo-induced degradation even in Ar gas. Despite its similarity to that of the bulk, the Raman spectrum of 2L revealed noticeable differences. First, some of the peaks have frequencies 0.5 % lower (A, B, E, G & K) or higher (C & L) than those of the bulk, while the others did not show such significant changes. Additionally, most of the peaks of 2L were much broader than those of the bulk (see Figure S14 and Table S2). As increasing thickness, the changes in the peak frequencies (Figure 4b) and line widths became smaller and the Raman spectra approached that of the bulk. The abrupt decrease in line widths between 2L and 3L can be attributed to intrinsic lifetime broadening or spectral inhomogeneity owing to substrate-induced structural rippling which occurs more significantly in thinner layers.[44] Further studies are required for differentiation between the two.

Among all the peaks observed, in particular, the A & B (C & L) peaks showed the largest frequency upshift (downshift). The hardening with increasing thickness can be understood from two oscillators coupled by a weak spring that corresponds to van der Waals interlayer interaction. Even a weak coupling essentially increases the restoring force leading to a mode hardening.[41, 45] Then the opposite softening is anomalous and attributed to the surface-layer effect as shown for 2D $MoS_2$ crystals.[46, 47] In essence, force constants (lengths) of intralayer bonds are larger (shorter) for the outermost layers than inner ones owing to lack of neighbors for the former. Because surface layers account for a larger fraction in thinner layers, the surface effect can influence lattice vibrations of single and few-layered 2D crystals more significantly. Quantitative understanding, however, requires theoretical investigation on the competition between the mode softening by the surface effect and the hardening by vdW interlayer coupling. It is also to be noted that thickness-dependent dielectric screening of interlayer Coulomb interaction plays a minor role.[46, 47] The thickness-dependent Raman spectral features can be readily employed in determining the thickness of a given $CrPS_4$ sample. Most of all, the lack of the Raman signal can serve as a fingerprint of 1L $CrPS_4$, which has an optical contrast sufficient for identification under an optical microscope, only slightly weaker than that of 2L (Figure S11). As shown in Figure 4c, the frequency difference between two peaks of opposite thickness-dependence is a robust function of thickness, which was confirmed with the three different excitation wavelengths. Overall, identification of 1L ~ 5L can be readily made based on the Raman results which can be further assisted by the optical contrast (Figure S11). While B and L peaks were selected for their large frequency changes and high intensities in Figure 4c, other pairs can be used instead. As shown in Figure 4d, the intensities of the major Raman peaks were proportional to the degree of absorption (at the excitation wavelength) that also scaled linearly with the thickness (see below). It is to be noted that the current Raman data represent the paramagnetic phase of $CrPS_4$ at room temperature. As shown for $FePS_3$,[20, 21] spin ordering below a critical temperature may influence the lattice dynamics and other optical properties of CrPS$_4$, and Raman spectroscopy will be very useful in characterizing its magnetic phase transition. Thus, the current Raman spectra will provide a firm basis for investigating magnetic properties of 2D CrPS$_4$ and their relation to various optical properties.

**Absorption and photoluminescence.** The electronic structure of 2D CrPS$_4$ near the Fermi level can be probed by optical absorption and emission. For the former, reflectance measurements were performed instead of transmission due to higher sensitivity.[48] For a very thin film (thickness $\ll \lambda$) supported on a thick transparent substrate with refractive index of $n_0$, absorbance (A) of the film can be determined from the fractional change in reflectance ($\delta_R$) as follows: $\delta_R = \frac{R-R_0}{R_0} = \frac{4}{n_0^2-1}A$, where R and $R_0$ are reflectance of the film and the bare substrate, respectively.[48] It is to be noted that the absorbance (A) refers to absorptance (fraction of incident intensity that is absorbed), not –log (transmittance). For these measurements, single and few layer CrPS$_4$ supported on quartz substrates were prepared. To identify very thin flakes and determine their thickness, the reflectance contrast was analyzed using the blue-channel data of the optical micrographs (Figure S15). As shown in the absorption spectra (Figure 5a), 1L is virtually transparent below 2.0 eV and exhibits a broad absorption peak centered at ~2.9 eV. With a closer look, however, another very weak absorption edge can be found near 1.5 eV. For 2L, the degree of absorption above 1.5 eV increased significantly and the high energy peak redshifted slightly to ~2.8 eV. With increasing thickness further, an absorption peak became noticeable at ~1.7 eV between the two absorption edges and the high energy peak gained intensity in proportion to the thickness (see the inset of Figure 5a). These two features are consistent with absorption spectra obtained in a transmission mode from bulk samples cleaved on a transparent adhesive tape (Figure S13).

Notably, the optical absorption spectra suggest that the electronic structure of 2D CrPS$_4$ is dependent on thickness but overall close to that of bulk.[13] Due to its ionic bonding nature, bulk CrPS$_4$ can be represented by Cr$^{3+}$[PS4]$^{3-}$ with each Cr atom located in a distorted octahedral interstice formed by 6 S atoms as schematically shown in Figure 5c. The bulk electronic structure near the Fermi level is dominated by 3d electrons of Cr and 3p nonbonding electrons of S.[25] According to the crystal field theory, 5d-orbitals of a metal ion in an octahedral ligand perturbation field split into t$_{2g}$ and e$_g$ orbitals with a splitting energy of $\Delta_o$ (Figure 5c; middle).[49] The ground electron configuration of Cr$^{3+}$ (t$_{2g}^3$) leads to a ground term of $^4$F, which splits further into three states of $^4$A$_{2g}$, $^4$T$_{2g}$, and $^4$T$_{1g}$ in the order of increasing energy (Figure 5c; right).[13] Many of Cr$^{3+}$ coordination compounds and solids exhibit two spin-allowed optical transitions, $^4$A$_{2g}$ → $^4$T$_{2g}$, and $^4$A$_{2g}$ → $^4$T$_{1g}$, which are located in the visible range of 1.6 ~ 2.2 and 2.2 ~ 3.2 eV, respectively.[50] Photoemission spectroscopy for bulk CrPS$_4$ showed that 3d-electrons of Cr and nonbonding 3p-electron of S are located 0.6 and 2.0 eV below the Fermi level, respectively.[25] The optical absorption spectrum of bulk CrPS$_4$ showed two edges at ~1.4 and ~2.2 eV, respectively. The former was attributed to the lowest d-d transition of d$^3$ chromium ion, which located the empty e$_g$ orbitals 0.8 eV above the Fermi level. This indicates that the octahedral S ligands impose relatively weak field[25] and thus generate small $\Delta_o$ which corresponds to the lowest d-d transition energy. The origin of the latter is unclear and can be the other d-d transition, ligand-to-metal charge transfer (LMCT: S 3p to Cr 3d) or a mixture of both.[13] Reflection electron energy loss spectroscopy (REELS) revealed two lowest d-d transitions at 2.0 and 2.5 eV, and two prominent LMCT peaks at 3.5 and 4.6 eV.[25] The difference in energy values and assignments may be due to large overlaps between the spectral peaks and unequal selection rules between the two methods.

The salient feature in the PL spectra of CrPS$_4$ (Figure 5b) is the emission peak at 1.31 ± 0.01 eV

slightly below the low-energy absorption edge. With decreasing thickness, its peak energy remained virtually constant and its intensity decreased rapidly in a nonlinear manner with respect to the thickness. In particular, 1L gave almost no signal in Figure 5b. Other samples of 1L generated a barely detectable signal but still much smaller than half of that for 2L. As shown in the inset of Figure 5b, the relative quantum yield $\Phi$ (defined as maximum PL intensity divided by absorbance at the excitation wavelength) decreased rapidly with decreasing thickness, which can be attributed to photodegradation as will be discussed below. The PL peak is attributed to spin-allowed d-d transition of Cr ($^4T_{2g} \rightarrow {}^4A_{2g}$) as found in various $Cr^{3+}$ coordination compounds and solids.[51-53] Its large Stokes shift of ~0.3 eV can be attributed to the configurational relaxation of the ligands.[52] Briefly, the equilibrium distance between $Cr^{3+}$ and ligands is different for the excited state $^4T_{2g}$ than the ground state $^4A_{2g}$, generating a substantial Franck-Codon offset along the configuration coordinate. Thus, the initial vertical transition from $^4A_{2g}$ leads to vibronic excitation of $^4T_{2g}$ and is followed by vibrational relaxation giving the Stokes shift as in molecular systems.

The above absorption and emission spectra revealed that optical transitions of 2D $CrPS_4$ in the visible and near-IR ranges resemble those of its bulk form and are dominated by d-d transitions of $Cr^{3+}$ ions under pseudo-octahedral crystal field perturbation. The slight peak energy changes in the absorption spectra can be attributed to the stacking or confinement effects.[54, 55] Despite the weak interlayer cohesion found in similar materials,[17] the presence of neighboring layers can affect the electronic structure of otherwise atomically thin 1L $CrPS_4$. In the crystal field model, next-nearest neighbors beyond the ligands in direct coordination affect $\Delta_o$ of the central metal ion significantly as comparatively shown in $CrF_3$ and $K_2NaCrF_6$.[56] We note that the thickness-dependent dielectric screening is also relevant. For example, the dielectric tensor of $MoS_2$ is highly anisotropic and its diagonal elements are larger in magnitude for bulk than 1L.[47] Assuming a similar trend, the inter-ionic interactions including the strength of the crystal field imposed by S ligands will be more screened in thicker $CrPS_4$, which may explain the observed shift. However, the fact that the absorption of the visible energy was linearly proportional to the thickness without significant change in the line shapes suggests that the overall electronic structure is largely unaffected by stacking in 2D $CrPS_4$.

It is also to be noted that the PL intensity is highly nonlinear and can be very weak for 1L, as summarized in the decreasing PL quantum yield for thinner layers. We attribute this to quenching by defects generated by excitation laser beam. Although the average power density of the excitation beam in the PL measurements was kept below 0.1 mW/μm$^2$, the gradual change was inevitable in the subsequent measurements in the Ar atmosphere. Zero Raman and negligible PL signals for 1L but still with finite absorption at the excitation wavelengths suggests that the thinnest $CrPS_4$ is unusually susceptible to photodegradation and other external perturbation. Since the early oxidation study of graphene,[57] similar thickness dependence has been reported for hydrogenation,[58] arylation,[59] charge transfer reactions,[60] etc. Other 2D crystals like $MoS_2$[61] and black phosphorus[62] also exhibited thickness-dependent oxidation behaviors. Considering the potential of TTMCs in 2D magnetism, the ambient stability and photoreactions of 2D $CrPS_4$ need to be clearly understood and requires systematic investigations.

**CONCLUSIONS**

We investigated structural and optical properties of single - and few-layer $CrPS_4$ synthesized by CVT method using TEM, AFM, optical microscopy, and Raman/reflectance/PL spectroscopies. Synthesized $CrPS_4$ showed single-crystallinity and its XRD data matched precisely with the theoretical prediction. By analyzing the preferred cleavage angles with TEM and OM, the crystallographic orientation of the exfoliated samples could be easily identified. The polarized microscopy revealed a strong in-plane optical birefringence, which can be also used in determining its lattice orientations. From the Raman measurements with three different laser wavelengths, certain peaks showed variation with thickness evolution, which can be used as an indicator of thickness. Absorption and PL measurements revealed that this material is a semiconductor with an optical band gap of $1.31 \pm 0.01$ eV, which is attributed to a d-d type transition of $Cr^{3+}$ ions. Due to its antiferromagnetic ordering below 36 K and semiconducting behavior, $CrPS_4$ has a high potential in 2D magnetism study and spintronic devices. The TEM and optical analyses in the current will provide an effective and accurate method to determine thickness and crystallographic orientation of 2D $CrPS_4$ in studying its novel material properties and applications. The observed strong structural and optical anisotropy suggests that its electronic and magnetic behaviors are also orientation-dependent and many intriguing properties are to be discovered for this and similar crystal systems.

## METHODS

**Synthesis process.** Single crystalline $CrPS_4$ was synthesized using CVT method. The mixture of powdered elements (99.5 % Cr, 99.99 % P and 99.5 % S with a molar ratio of 1:1:4; Sigma-Aldrich) was vacuum-sealed into a quartz ampoule. This ampoule was placed into a tubular furnace, which has a temperature gradient, and the source side was in the high-temperature zone. The furnace temperature ramped to 700 °C at the rate of 1 °C/min and was kept at 700 °C for a week. After that, the ampoule was slowly cooled down to room temperature and broken to collect the synthesized crystals.

**Characterizations.** The flakes of 40 mg were put in 100 ml of N-methyl-2-pyrrolidone (NMP) and dispersed using a sonication bath for 4 h. To obtain the powdered $CrPS_4$, the solution was filtered through a porous membrane. The filtered powder was kept in a vacuum oven to remove the solvent. XRD patterns of the powder and single crystal bulk material were obtained by a diffractometer (D8 Advance, Bruker Corporation) using Cu K-alpha radiation (1.5406 nm) at room temperature. Mechanically exfoliated $CrPS_4$ on $SiO_2$/Si substrate was picked up by thin polycarbonate (PC) film and transferred to a TEM grid. Residual PC film was removed by chloroform solvent. The focused ion beam (FIB) technique was employed to fabricate the sample for the cross-sectional view. Also, exfoliated $CrPS_4$ was FIB-milled until it is electron-transparent. Then, the thin lamella was removed from its trench and transferred to a TEM grid. The atomically resolved TEM and STEM image were carried out using an image Cs-corrected Titan at 80 kV and probe Cs-corrected JEM-2100F at 200 kV.

**Optical spectroscopies.** Thin $CrPS_4$ samples (1L ~ 10 nm) for the optical measurements were prepared by mechanical exfoliation of the CVT-grown crystals onto $SiO_2$/Si or amorphous quartz substrates. The home-built setup for the micro-Raman and PL measurements has been described elsewhere.[63, 64] Briefly, the output of visible lasers (two solid state lasers for 457.9 and 514.3 nm; one He-Ne gas laser for 632.8 nm) was focused onto a 1-μm diameter spot on a sample using an objective lens (40x, numerical aperture = 0.60). The back-scattered signal was collected by the same objective and fed into a Czerny-Turner spectrometer (focal length = 300 mm) equipped with a charge-coupled detector (CCD) cooled with liquid nitrogen. The spectral resolution defined by the FWHM of the Rayleigh peak was 3.0 $cm^{-1}$ and 1.5 meV for the Raman and PL measurements, respectively. For the reflectance measurements, a similar microscope-based spectrometer with a thermoelectically-cooled CCD was used with a broadband collimated light beam from a tungsten halogen/deuterium lamp. To avoid the photodegradation during the Raman and PL measurements, samples were placed in an optical cell filled with high-purity Ar gas (flow rate = 1.0 L/min). Reflectance measurements were carried out in the ambient conditions.

**ACKNOWLEDGMENT**

This work was supported by the Institute for Information & Communications Technology Promotion (IITP) grant (B0117-16-1003), the Center for Advanced Soft-Electronics funded by the Ministry of Science, ICT and Future Planning as Global Frontier Project (NRF-2011-0031630) and by the National Research Foundation of Korea (NRF-2012R1A3A2048816, 2015R1A2A1A15052078, 2015R1A2A2A01006992, and 2016R1A2B4012931).

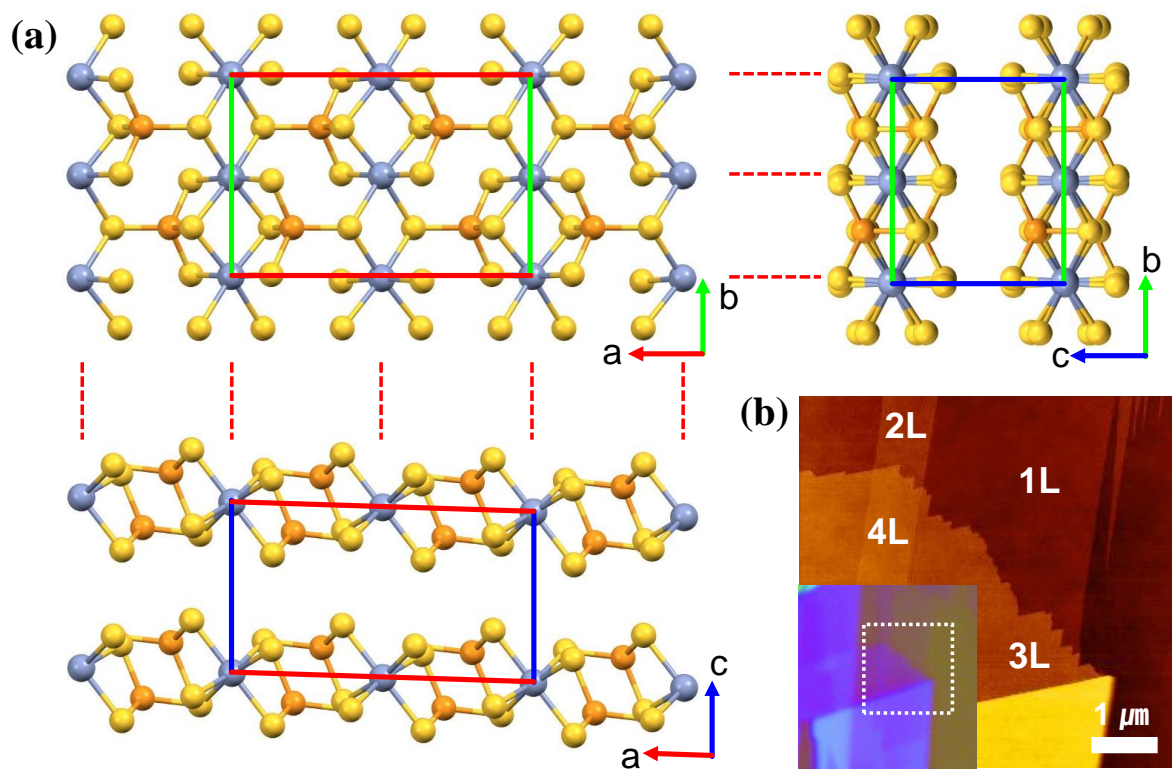

**Figure 1.** Crystal lattice structure and exfoliated sample of CrPS$_4$. (a) Third angle orthographic projection of layered CrPS$_4$ crystal. Steelblue, orange, and yellow spheres represent Cr, P and S atoms, respectively. The red, green and, blue lines show lattice unit cell, and its lattice constants are a= 10.871, b= 7.254, and c= 6.140 Å, respectively. Each arrow indicates axis directions. (b) AFM image of exfoliated 1 to 4 layers of CrPS$_4$. The inset is the optical microscope image, in which the square box corresponds to the AFM scan area.

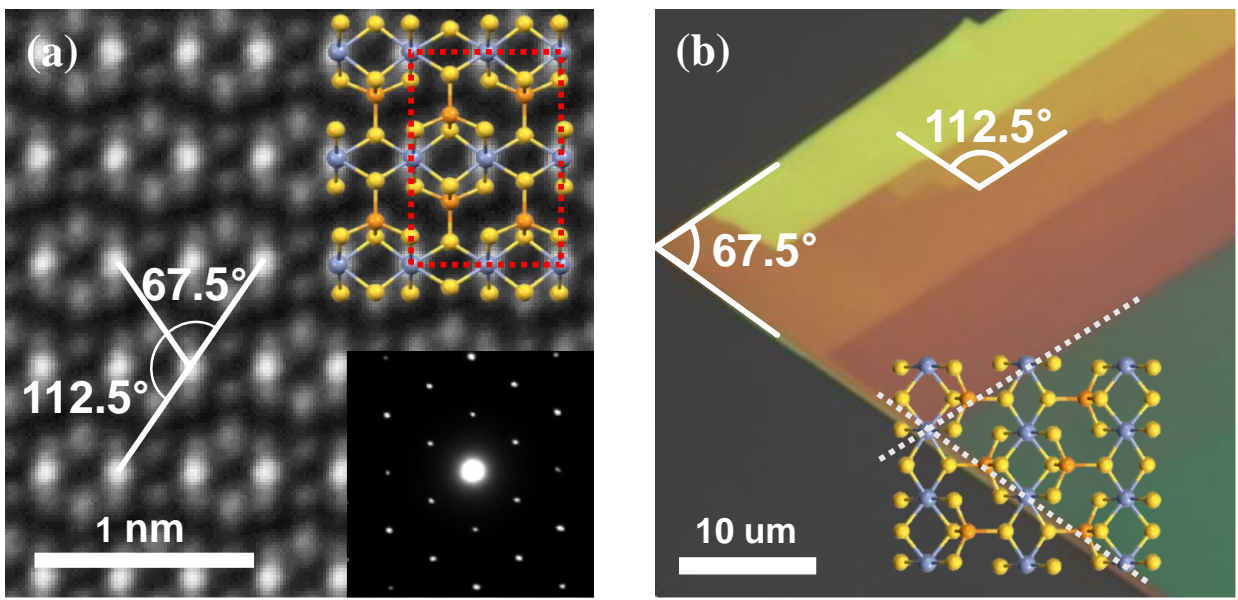

**Figure 2.** Structural analysis on CrPS$_4$. (a) STEM image in high magnification clearly reveals atom alignment of a few layered sample. The inset shows its SAED pattern and single crystallinity. An overlaid structural model reveals atom arrangement. (b) Thin sample exfoliated on a silicon substrate with 300 nm-thick SiO$_2$. Straight edges formed preferential angles as marked in the image. The angles between the diagonal direction of Cr atoms agreed well the angles in (a).

Figure

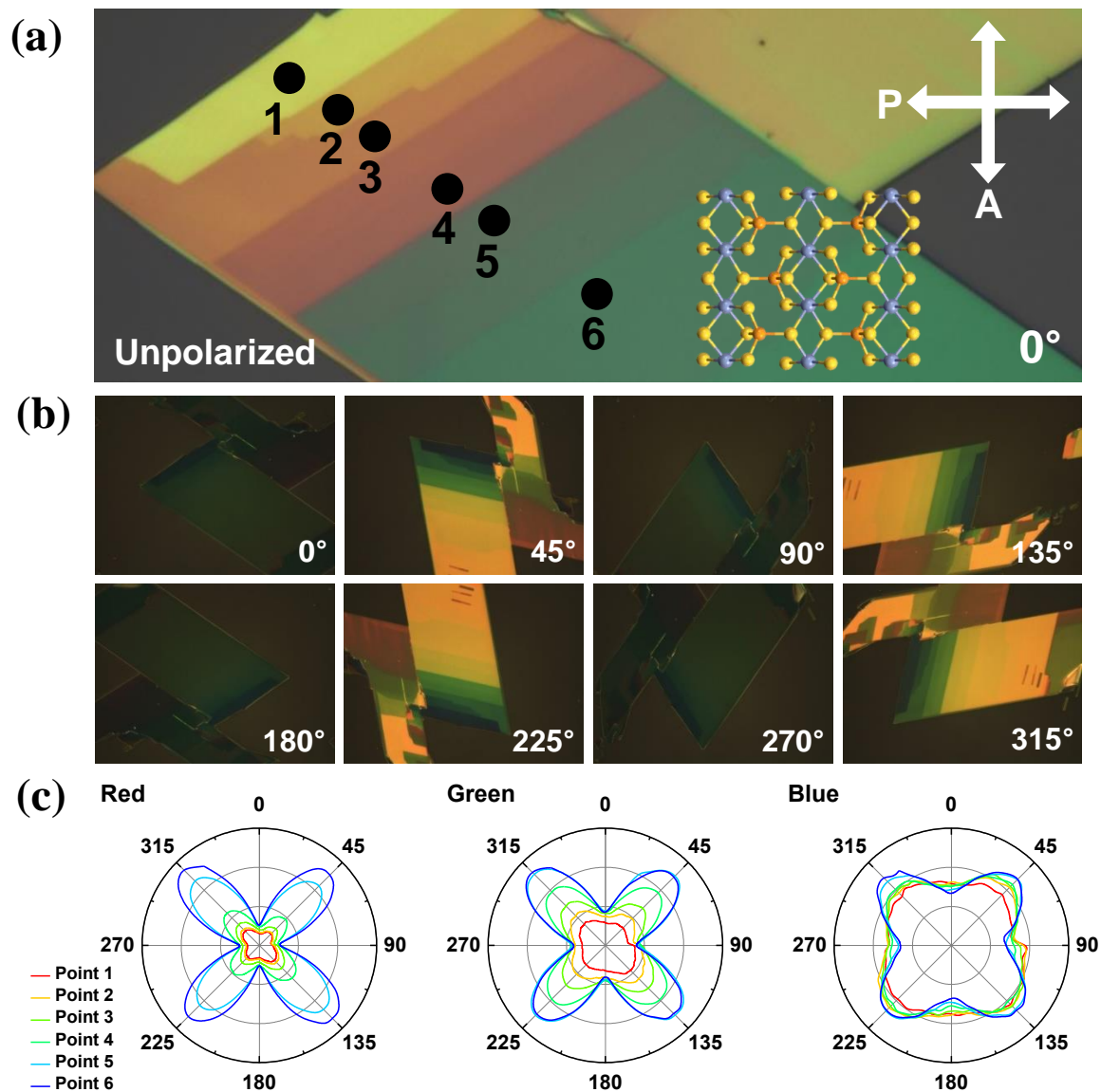

**Figure 3.** Optical anisotropy observed by the polarized microscopy. (a) Unpolarized and (b) polarized optical micrographs of exfoliated CrPS$_4$. The arrows represent the directions of the fixed polarizer (P) and analyzer (A). The sample was rotated clockwise from initial 0º to angles shown in optical images. (c) Angular dependence of RGB values that were collected from six positions with different thicknesses as marked in (a).

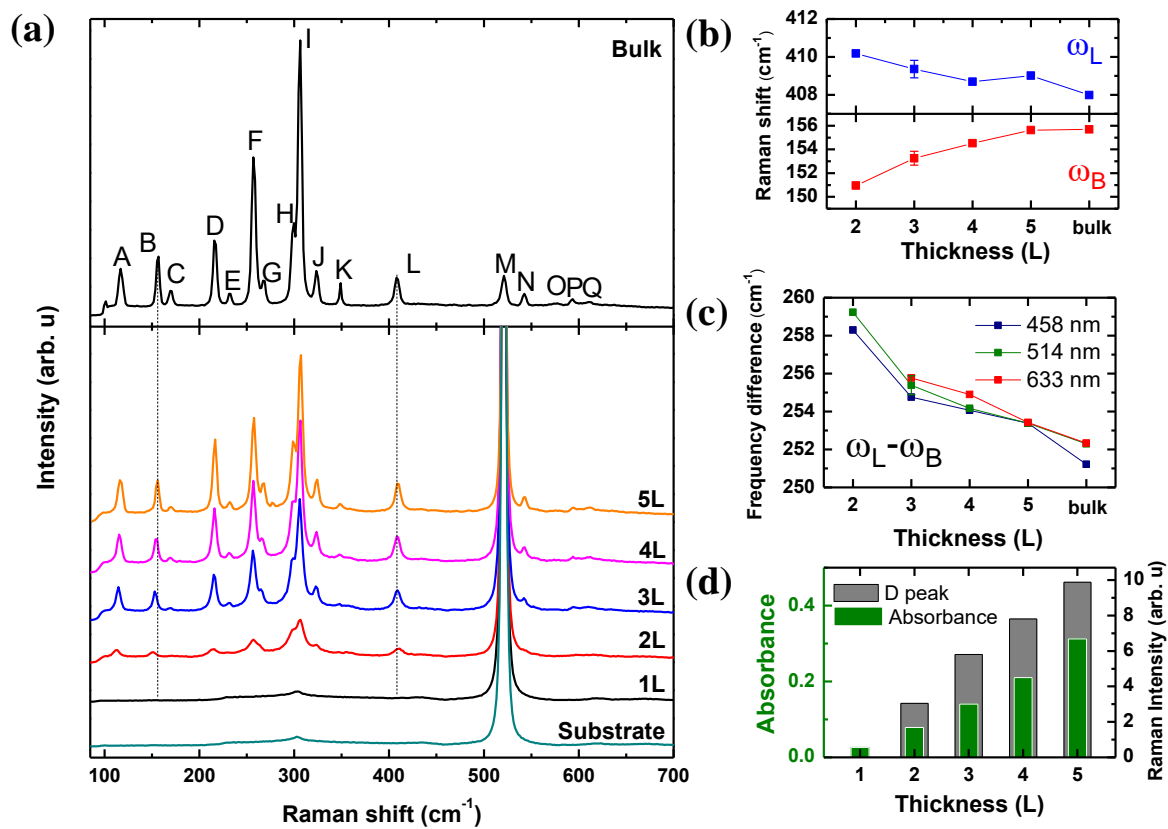

**Figure 4.** Dependence of lattice vibrations of 2D CrPS$_4$ on its thickness. (a) Raman spectra of 1L ~ 5L CrPS$_4$ in comparison with that of its bulk form (thickness = ~40 nm). The signal at 520 cm$^{-1}$ is mostly due to Si rather than M peak for thin layers. (b) Peak frequencies of the two representative Raman modes, upshifting $\omega_B$ and downshifting $\omega_L$ with increasing thickness. (c) Frequency differences between the two modes ($\omega_L$-$\omega_B$) as a function of thickness, obtained for three different excitation wavelengths: 458, 514, and 633 nm. (d) Comparison between the intensity of D peak and optical absorption given as a function of thickness.

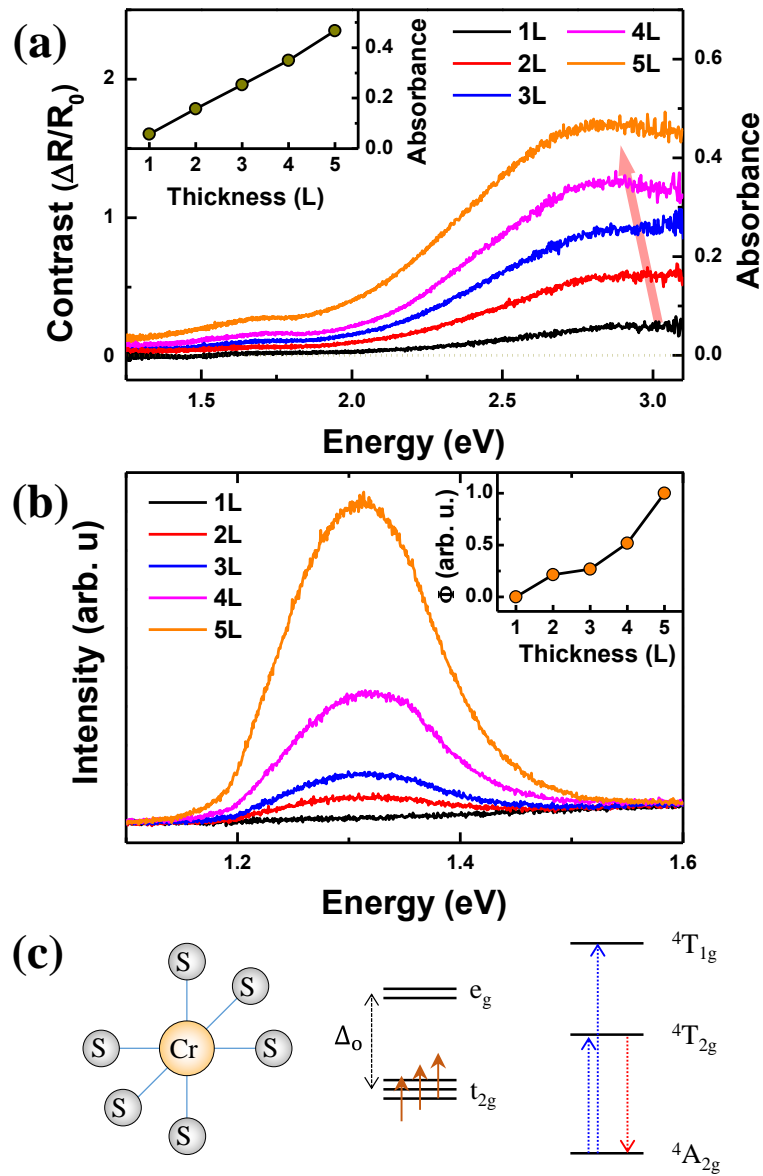

**Figure 5.** Electronic structure of 2D CrPS$_4$ revealed by optical measurements. (a) Reflectance contrast (left ordinate) and absorptance (right ordinate) spectra of 1L ~ 5L CrPS$_4$. The inset reveals a quasi-linear relation between maximum absorption averaged over 2.8 ~ 2.9 eV and thickness. (b) PL spectra of 1L ~ 5L CrPS$_4$. The inset shows that the relative quantum yield (Φ) decreases rapidly with decreasing thickness. (c) Schematic diagrams of geometrical and electronic

structure responsible for the d-d transition: (left) Cr atom bonded to six S ligands in a distorted octahedral structure; (middle) ground-state electron configuration of $Cr^{3+}$ ions under octahedral crystal field which splits five degenerate d-orbitals into $t_{2g}$ and $e_g$ orbitals by $\Delta_o$; (right) three lowest states for the ground-state configuration of $Cr^{3+}$, $(t_{2g})^3(e_g)^0$, with allowed optical transitions denoted by arrows.